\documentstyle[12pt]{article}

\begin{document}
\begin{center}
{\Large \bf The degree scale feature in the CMB spectrum
in the fractal universe }
\bigskip

{\large D.L.~Khokhlov}
\smallskip

{\it Sumy State University, R.-Korsakov St. 2, \\
Sumy 244007, Ukraine\\
E-mail: khokhlov@cafe.sumy.ua}
\end{center}

\begin{abstract}
The position of the degree scale feature in the CMB spectrum
is determined within the framework of the fractal universe
with a power index of 2.
\end{abstract}

The presence of any feature in the cosmic microwave background (CMB)
anisotropy spectrum whose physical scale is known provides us with
the ability to perform the angular diameter distance test~\cite{Hu}.
Any feature progects as an anisotropy onto an angular scale
associated with multipole
\begin{equation}
\ell_{eff}=kd
\label{eq:mul}
\end{equation}
where $k$ is the size of the feature in k-space, $d$ is
the angular diameter distance to the feature.
Anisotropy measurements on degree scales pin down the feature in the
CMB spectrum. The position of the feature is
$\ell_{eff}=263$~\cite{Ha},
$\ell_{eff}=260$~\cite{Li}.
In the Friedmann universe, the degree scale feature is considered
to arise due to acoustic
oscillations in the photon-baryon fluid at last scattering.
The feature represents the sound horizon when
the universe recombines.

The potential fluctuation $\Delta M/M$ is a relation of the
scales. The mass $M$ defines the scale of homogeneity.
The mass $\Delta M$ defines the scale of fluctuations.
The scale of homogeneity can be specified with the distance $r$ covered
by the CMB photon.
The distance $r$ gives the mass of radiation $M$ restricted within $r$.
Then the size of the potential fluctuation
$\Delta r$ is defined via the value $\Delta M/M$.
The size of the potential fluctuation
$\Delta r$ represents the feature in the CMB spectrum,
with the value $r$ being the distance to the feature.

Determine the above feature in the CMB spectrum
within the framework of the universe
with the linear evolution law~\cite{Kh1}.
The model is based on the premise that
the evolution of the universe is not
defined by the matter but is a~priori specified.
The scale factor of the universe is a linear function of time
\begin{equation}
a=ct.\label{eq:g1}
\end{equation}
The scale of mass is a linear function of the scale factor
\begin{equation}
M\sim a.
\label{eq:M}
\end{equation}
This leads to that the universe has the fractal structure with
a power index of 2~\cite{Kh2}.
In the fractal universe, the relation of the masses within radii
$R_1$ and $R_2$ is given by
\begin{equation}
\frac{M_{1}(<R_{1})}{M_{2}(<R_{2})}=
\left(\frac{R_{1}}{R_{2}}\right)^2.
\label{eq:mr}
\end{equation}
Then the size of the potential fluctuation is given by
\begin{equation}
\frac{\Delta r}{r}=\left(\frac{\Delta M}{M}\right)^{1/2}.
\label{eq:dr}
\end{equation}
In view of this, the multipole in the CMB spectrum is
given by
\begin{equation}
\ell_{eff}=\left(\frac{\Delta r}{r}\right)^{-1}=
\left(\frac{\Delta M}{M}\right)^{-1/2}.
\label{eq:ell}
\end{equation}
According to~\cite{Kh3}, the value of potential fluctuations
of radiation is $\Delta M/M=1.49 \times 10^{-5}$.
From this calculations yield $\ell_{eff}=259$.

\end{document}